\documentclass[conference]{IEEEtran}
\IEEEoverridecommandlockouts
\usepackage{cite}
\usepackage[english]{babel}
\usepackage{amsmath,amssymb,amsfonts}
\usepackage{algorithmic}
\usepackage{graphicx}
\usepackage{textcomp}
\usepackage{xcolor}
\usepackage{ctable}
\usepackage{verbatim}
\usepackage{enumitem,hyperref}
\usepackage{array}
\usepackage{float}
\usepackage[breakable]{tcolorbox}
\usepackage{tablefootnote}
\def\BibTeX{{\rm B\kern-.05em{\sc i\kern-.025em b}\kern-.08em
    T\kern-.1667em\lower.7ex\hbox{E}\kern-.125emX}}
\begin{document}

\title{Practices, Challenges, and Opportunities When Inferring Requirements From Regulations in the FinTech Sector - An Industrial Study\\
}

\author{
    \IEEEauthorblockN{1\textsuperscript{st} Anonymous Author}
    \IEEEauthorblockA{\textit{Anonymous University)}\\
    City, Country \\
    anonymous.author@uni.com}
}

\author{\IEEEauthorblockN{1\textsuperscript{st} Parisa Elahidoost}
\IEEEauthorblockA{\textit{fortiss GmbH} \\
Munich, Germany \\
elahidoost@fortiss.org}
\and
\IEEEauthorblockN{2\textsuperscript{nd} Daniel Mendez}
\IEEEauthorblockA{\textit{Blekinge Institute of Technolog and fortiss GmbH} \\
Karlskrona, Sweden \\
daniel.mendez@bth.se}
\and
\IEEEauthorblockN{3\textsuperscript{rd} Michael Unterkalmsteiner}
\IEEEauthorblockA{\textit{Blekinge Institute of Technology} \\
Karlskrona, Sweden \\
michael.unterkalmsteiner@bth.se}
\and
\IEEEauthorblockN{4\textsuperscript{th} Jannik Fischbach}
\IEEEauthorblockA{
\textit{Netlight Consulting GmbH and fortiss GmbH}\\
Munich, Germany \\
jannik.fischbach@netlight.com}
\and
\IEEEauthorblockN{5\textsuperscript{th} Christian Feiler}
\IEEEauthorblockA{\textit{itestra GmbH} \\
Munich, Germany \\
feiler@itestra.de}
\and
\IEEEauthorblockN{6\textsuperscript{th} Jonathan Streit}
\IEEEauthorblockA{\textit{itestra GmbH} \\
Munich, Germany \\
streit@itestra.de}
}

\maketitle

\begin{abstract}
\textbf{[Context and motivation]}: Understanding and interpreting regulatory norms and inferring software requirements from them is a critical step towards regulatory compliance, a matter of significant importance in various industrial sectors. 

\textbf{[Question/ problem]}: 
However, interpreting regulations still largely depends on individual legal expertise and experience within the respective domain, with little to no systematic methodologies and supportive tools to guide this practice. In fact, research in this area is too often detached from practitioners' experiences, rendering the proposed solutions not transferable to industrial practice. As we argue, one reason is that we still lack a profound understanding of industry- and domain-specific practices and challenges. 

\textbf{[Principal ideas/ results]}:  We aim to close this gap and provide such an investigation at the example of the banking and insurance domain. We conduct an industrial multi-case study as part of a long-term academia-industry collaboration with a medium-sized software development and renovation company. We explore contemporary industrial practices and challenges when inferring requirements from regulations to support more problem-driven research. Our study investigates the complexities of requirement engineering in regulatory contexts, pinpointing various issues and discussing them in detail. We highlight the gathered insights and the practical challenges encountered and suggest avenues for future research. 

\textbf{[Contribution]}: Our contribution is a comprehensive case study focused on the FinTech domain, offering a detailed understanding of the specific needs within this sector. We have identified key practices for managing regulatory requirements in software development, and have pinpointed several challenges. We conclude by offering a set of recommendations for future problem-driven research directions.
\end{abstract}


\section{Introduction}
Software-intensive systems, particularly within the rapidly evolving FinTech sector, must comply with regulatory artifacts such as laws, policies, mandates, and guidelines. A first step towards compliance is ensuring that the software requirements elaborated in software development projects properly reflect those regulatory artifacts. This task, however, is already error-prone and labor-intensive~\cite{b14} and too often dependent on legal and domain experts who might, in turn, not always be available to the engineering teams. 

To enhance the capability of engineering teams in dealing with regulatory requirements, there is a pressing need for systematic methodologies and supportive tools tailored to the respective domain. Such advancements necessitate a thorough understanding of the current practices and challenges specific to the sector, thereby enabling research and development efforts to be more problem-driven and relevant to industry needs. Motivated by this, our study embarks on a multi-case analysis at \textit{itestra}, a medium-sized company with significant experience in software development and renovation projects within highly regulated domains. By focusing on the banking and insurance aspects of the FinTech industry, we aim to focus on the specific practices and hurdles encountered therein, contributing to a more targeted body of research that addresses the needs of this domain.

This paper reports on our investigation into how regulatory requirements are integrated into software development processes at \textit{itestra}, exploring the particular challenges of the FinTech sector and suggesting paths for future work. Our study is part of a long-term academia-industry collaboration and aims to support our research and development activities in a more industrially relevant, problem-driven manner. Modern software development is shaped by the needs and problems identified by various stakeholders, which are articulated through requirements that guide the development of software products. The focal point of this study is to gain a deeper understanding of how practitioners manage the incorporation of regulatory requirements into their engineering processes, pinpointing the existing challenges and exploring and providing insights for potential improvements that are not only relevant but also actionable for practitioners within this specialized field.

The remainder of this paper is organized as follows. We present a brief overview of the existing literature and background terminology in Section~\ref{sec:BackgroundRelWork}. We introduce our research goals and methodology in Section~\ref{sec:ResearchMethods}. In Section~\ref{sec:Results}, we present the results, structured along our research question and critically discuss them in Section~\ref{sec:Discussion}, before concluding our work with Section~\ref{sec:Conclusion}. 

\section{Background and Related Work}
\label{sec:BackgroundRelWork}
\subsection{Terminology}
This section provides an overview of the most relevant terms and concepts used in the context of our study.

\textit{Regulatory compliance} is the act of ensuring adherence of an organization, process or (software) product to laws, guidelines, specifications and regulations~\cite{b15}. 

\textit{Regulatory Requirements compliance} is the degree of adherence of functional and non-functional requirements and constraints to criteria laid out in regulations. 

\textit{Regulation} is any official document that is a source of public, general, obligatory norms issued by regulators~\cite{b16}. We also consider standards as a type of regulation.

\textit{FinTech} is defined as a new financial industry that leverages technology to enhance financial activities. Typically, FinTech is used when the financial sector makes use of the availability of ubiquitous communication, specifically via the Internet and automated information processing~\cite{b24}.

\subsection{Related Work}
To collect related approaches, we are analyzing available secondary studies on the topic of regulatory compliance. 

In 2022, Ardila et al.~\cite{b20} conducted a systematic review on compliance checking in software engineering, identifying 41 studies. They analyzed methods for automatic compliance-checking, its impacts, and challenges, categorizing five main challenges, including modeling language use, normative requirements suitability, generic method development, automation enhancement, and application issues, without directly linking them to specific study findings. 

Mubarkoot et al.~\cite{b22} performed a systematic review of software compliance requirements, highlighting human and technology challenges and suggesting policies as solutions. Despite categorizing these into technological and human-related issues, the study's broad classifications fall short of the detail necessary for practical application in engineering regulatory compliance. 

Usman et al.~\cite{b5} conducted a study closely related to ours, exploring compliance requirements in large-scale software development via a case study focused on a single organization. The study provides insights into compliance strategies and industry-specific challenges but is limited by its single-case scope, potentially affecting the findings' applicability to various contexts. This suggests the need for wider research with multiple case studies for a fuller understanding of compliance in software development. 

Mustapha et al.~\cite{b17} investigates managing compliance requirements in business processes, examining methodologies and practices for adherence to standards and regulations. They identify research trends, challenges, and future areas for exploration, providing insight into the complexities of incorporating compliance into business process management and summarizing the field's current research state. 

Akhigbe et al.~\cite{b18} reviewed 103 studies on goal-oriented and non-goal-oriented modeling for legal compliance, identifying benefits and limitations and focusing on goal modeling challenges, notably the limited use of goal models for comparing models from legal texts. 

Hashmi et al.~\cite{b19} studied business process compliance practices, identifying limitations and future research opportunities. Their survey consolidated challenges in ensuring compliance, which laid the groundwork for improved methods in the field. 

Naira et al.~\cite{b21} developed a taxonomy from 218 studies to categorize safety assurance evidence and artifacts, addressing challenges frequently addressed in academic research related to safety evidence to enhance application across domains.

This study aims to fill the existing knowledge gap by investigating practices and challenges in handling regulatory compliance requirements in software engineering. Previous research, including systematic reviews and case studies, has highlighted various challenges and proposed compliance checking and management methods. However, these studies often lack detailed classification of challenges or focus on specific compliance aspects, such as modeling methods or evidence management. By conducting a multi-case study in a real-world setting, our research seeks to provide a comprehensive understanding of how regulatory requirements are integrated into software development processes, identify common challenges, and explore the potential for improvements.

\section{Research Methodology}
\label{sec:ResearchMethods}

This study was conducted to gain in-depth insights into the current state of the practices and challenges faced by practitioners when handling requirements derived from regulatory sources at the case company. Given the exploratory nature of this investigation, we adopted a multiple-case study approach~\cite{b1} with four distinct cases. We collected the data through a triad of methods: conducting interviews, facilitating group discussions, and examining corporate documentation. 

\subsection{Goal} \label{AA}
The primary aim of this study is to explore software engineering practices and challenges when working with requirements that emerge from regulatory sources. Additionally, this study seeks to identify the constraints and opportunities for enhancement in dealing with such requirements within a medium-sized enterprise operating in the FinTech sector, thereby establishing the groundwork for the future formulation of a sensible approach that accommodates these limitations. We address our study goal through the following research questions:

\begin{enumerate}[label=\textbf{$RQ$\arabic*}]
    \item \label{rq1} \textbf{What are the engineering practices when working with requirements with regulations as a source?}
    
    Motivation: To explore how software engineering practices adapt to regulatory requirements by examining the methods used to interpret and incorporate these regulations into development processes, focusing on the activities, roles, and procedures.
    \item \label{rq2} \textbf{What are the associated challenges encountered by software engineering teams when working with requirements stemming from regulations?}
    
    Motivation: To identify the challenges and issues faced by practitioners in medium-sized companies 
    as they integrate regulatory requirements into software engineering practices, aiming to understand the complexities of complying with legal and regulatory standards in modern software development.
    \item \label{rq3} \textbf{How can tool support ease engineering activities, highlighted by the identified challenges?}
    
    Motivation: To explore engineering practices in software development impacted by the challenges of integrating regulatory requirements. The aim is to evaluate how tool support can ease these challenges, thereby improving compliance processes and efficiency and reducing the regulatory burden on development teams.
\end{enumerate}

\subsection{Context Description}
In this study, our case company is \textit{itestra}, a company specializing in developing and renovating large tailor-made business information systems.  Operating in 12 locations across four countries, \textit{itestra} employs over 130 computer scientists to build custom software for global corporations and successful mid-sized businesses. Further details about the selected cases are in section \ref{subsub:PS}. 

\subsection{Data Collection and Analysis}
\begin{figure}
 \centering
  \includegraphics[width=\columnwidth]
    {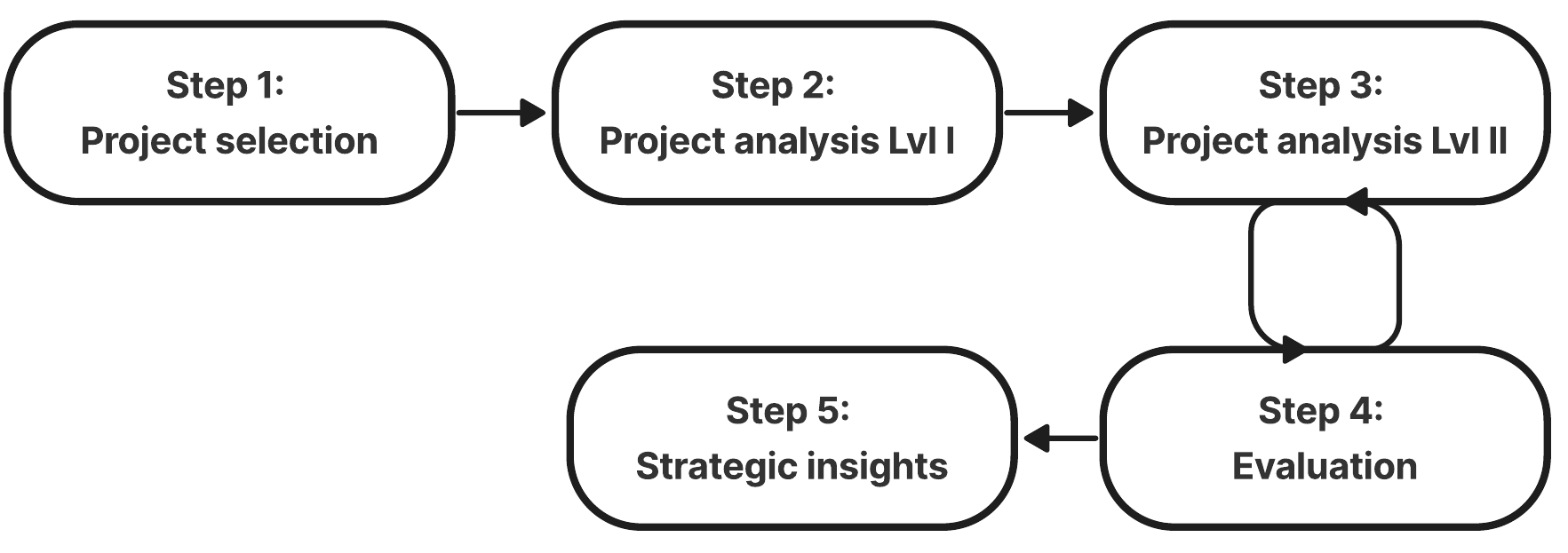} 
    \caption{Visualization of the research method steps} 
    \label{RM}
\end{figure}

The process of our study, visualized in Fig.~\ref{RM}, was structured into five distinct steps. The initial step involved a selection process through a workshop and group discussion, including the CEO and senior roles with a comprehensive overview of all company projects aimed at identifying relevant projects and roles. Group discussions played a crucial role in this process, as they enabled the collection and selection of data through interactive dialogues involving numerous participants. Following the initial group discussion, we pinpointed projects within highly regulated sectors and identified key roles for data collection, streamlining our focus towards gathering pertinent information effectively. This was followed by the second step, project analysis and data elicitation, achieved via semi-structured interviews and document analysis. Semi-structured interviews were employed for data collection in this study, as they offer the flexibility for improvisation and in-depth exploration of the subject matter~\cite{b1}. The third step encompassed the construction of a map with identified artifacts and roles, which served as a foundational tool for the subsequent step. The fourth step, evaluation, engaged all interview participants in the process of collaborative assessment. This step consisted of the presentation and explanation of the artifact map and the identification of errors by the interviewees. We adopted the artifact map construction introduced in the paper by Unterkalmsteiner et al.~\cite{b2}. During this phase, interviewees were encouraged to spend a few minutes with the map to pinpoint any unclear artifacts, roles, or incorrect relationships between artifacts. The fifth and final step of our study's process entailed synthesizing the rich insights acquired from prior interactive phases into actionable recommendations to enhance the engineering process. Steps 2 to 4 were systematically carried out per selected project, with each step following the previous one in order, except for steps 3 and 4, which were part of an iterative process, allowing for incremental refinement and validation.

\subsubsection{Step 1 - Project Selection} \label{subsub:PS}
The study commenced with an initial open-ended group discussion, including a diverse mix of researchers and practitioners. This discussion aimed to pinpoint relevant projects and roles. Participants in this session included two researchers, the managing director, a sales director, and a senior analyst and principal engineer. \textit{itestra} has multiple customers and projects in different regulatory domains. The regulations in different domains have different levels of complexity and granularity. As shown in Table~\ref{tab1}, we selected four projects as our cases, which were randomly assigned index names from A to D to remove traceability back to the case company. To ensure comparability between the projects with respect to practices and challenges, they all are, in the wider sense, related to the banking and insurance domain. Notably, these projects have an duration extending beyond ten years. This longevity underscores regulatory compliance's complexity and evolving nature in software development. Such extended timelines highlight the need to adapt to regulatory changes and maintain compliance.

\begin{table}
\caption{Project Overview}
\begin{tabular}{lp{1cm}p{1cm}p{4cm}}
\toprule
& \textbf{Domain} & \textbf{Duration} & \textbf{Examples of Relevant Regulations} \\ \midrule
\textbf{Project A} & Banking & 2013 - present & \href{https://www.gesetze-im-internet.de/invstg_2018/BJNR173010016.html}{Investment Tax Act}  
\& \href{https://www.bzst.de/DE/Unternehmen/Intern_Informationsaustausch/CommonReportingStandard/Handbuecher_ab_2021/handbuecher_node.html#js-toc-entry2}{Common Reporting Standards} \\ 
\textbf{Project B} & Banking & 2013 - present & \href{https://www.irs.gov/businesses/corporations/foreign-account-tax-compliance-act-fatca}{FACTA}, \href{https://www.bzst.de/SharedDocs/Downloads/DE/CRS/KHB/crs_khb_teil_2_ab_20210101.pdf?__blob=publicationFile&v=15}{Communications Handbook}
\& \href{http://www.gesetze-im-internet.de/fkaustg/index.html}{Law on the automatic exchange of information about financial accounts in tax matters} \\ 
\textbf{Project C} & Insurance & 2014 - present & \href{https://web.archive.org/web/20190902072002/https://www.bundesfinanzministerium.de/Content/DE/Downloads/BMF_Schreiben/Weitere_Steuerthemen/Abgabenordnung/2014-11-14-GoBD.pdf?__blob=publicationFile&v=4}{GoBD} \& \href{https://www.gdv.de/resource/blob/90408/c391b1dd04b41448fdb99918ce6d03bf/download-code-of-conduct-data.pdf}{Code of conduct by insurance companies}\\ 
\textbf{Project D} & Public service & 2016 - present & \href{https://www.gkv-datenaustausch.de/arbeitgeber/deuev/gemeinsame_rundschreiben/gemeinsame_rundschreiben.jsp}{Data exchange with social insurance} \\ \bottomrule
\end{tabular}
\label{tab1}
\end{table}

\subsubsection{Step 2 - Project Analysis Level I}
In the second step, we conducted semi-structured interviews using a guide developed by the first author and independently reviewed by the second author. The interview guide (part of the disclosed protocol) contains seven sections: a preface outlining the study's objectives, a disclaimer, an inquiry into the participants' backgrounds, an examination of engineering practices, an assessment of challenges faced, an exploration of potential enhancements, and a concluding segment requesting access to relevant project documentation and artifacts. The interview questions can be found at \href{https://zenodo.org/uploads/10640987}{zenodo}.   
The interviews, conducted in May 2023 and lasting between 30 to 40 minutes each, involved the most senior project personnel from each project, resulting in a total of four in-depth interviews. This approach ensured that we gathered insights directly from those with extensive experience and knowledge of the projects' complexities and challenges.
All interviews were digitally recorded and transcribed verbatim with participants' consent for the recording and anonymous citation in this paper, ensuring accuracy. Furthermore, the interviewer explained the purpose of the interview and underscored the voluntary nature of the engagement and the participants' autonomy to withdraw at any time. Complementary to the interviews, the company's documentation, such as requirement specifications, was examined to augment the understanding of the projects' complexities.

\subsubsection{Step 3 - Project Analysis Level II}
\begin{figure}
 \centering
  \includegraphics[width=\columnwidth]
    {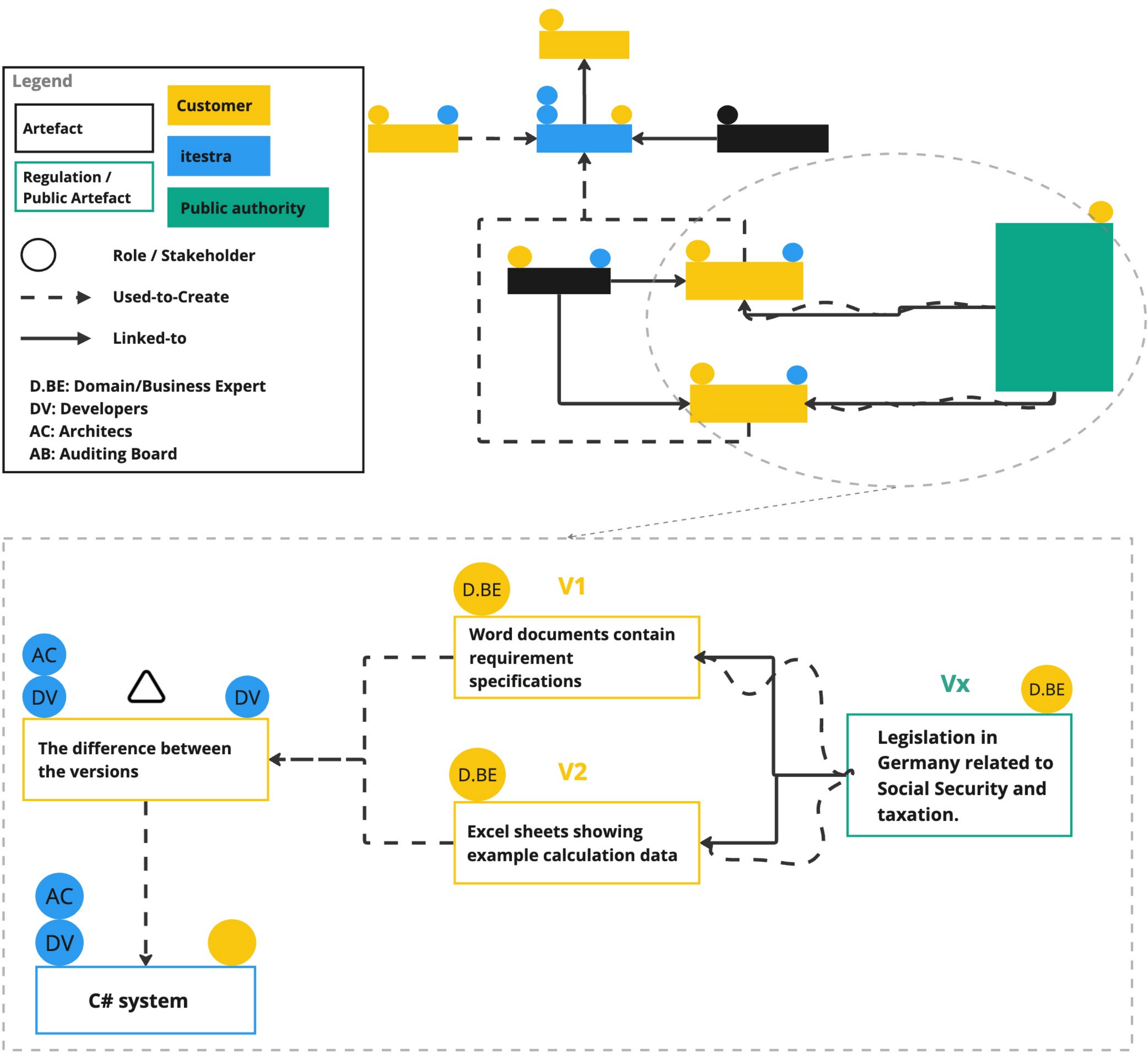} 
    \caption{Exemplary artifact map of project D (anonymized)} 
    \label{map}
\end{figure}

We examined the gathered qualitative data to identify, analyze, and report patterns and themes within the data. The analysis of the cases was conducted individually; however, a cross-case analysis was also partially undertaken during synchronization workshops and group discussions involving the CEO. This approach allowed for a detailed understanding of each case while also facilitating the identification of common themes and insights across all projects. Initially, audio recordings were transcribed word-for-word. Next, we created a general start list for clustering according to the research questions and context information. The start list acted as a preliminary theme to group the raw data according to the general domain instead of content-specific, enabling inductive coding. Subsequently, we crafted visual representations of roles and artifacts to synthesize the themes, which aided in simplifying complex information and facilitating communication with participants. These visualizations also served as a foundational tool for collaborative identification of challenges. Figure~\ref{map} illustrates parts of the artifact map generated for project D.
This includes using distinct shapes to signify artifacts and their creators or users, and varied line styles to represent different types of relationships between artifacts. Color coding was strategically employed to denote the origin of data points, enhancing the map's clarity. The employment of artifact maps facilitated a comprehensive overview of the information flows within and across projects, enabling the pinpointing of specific process or artifact areas where challenges were prevalent. 

\subsubsection{Step 4 - Evaluation}
To ensure accuracy and mitigate interpretation risks, interviewees were involved in the validation process, conducting a detailed review of data clusters and artifact maps, highlighting specific challenges and opportunities for improvement. This review was done per project and focused on verifying the correctness of the data and identifying any inaccuracies or errors. This step included refining roles based on their feedback, with one interviewee suggesting a specific update. This collaborative verification approach with stakeholders further corroborated that we recorded the data accurately and further synthesized and interpreted it correctly. In addition, the finalized artifact maps, along with the pinpointed challenges and identified opportunities for enhancements, for all cases, were also presented, discussed, and agreed upon in one of the group discussions.

\subsubsection{Step 5 - Strategic Insights and Action Plan}
This step summarises our research into actionable insights and strategies, ensuring they are effectively communicated to a broad audience. The presentation aims to be concise yet sufficiently detailed to serve as a resource for those not involved in the assessment or the studied projects. The purpose is to distil our research findings into practical insights and actionable strategies and facilitate cross-analysis, ensuring these insights are communicated to a wide audience in the organization. This approach is designed to be succinct and informative, providing valuable guidance to those not directly involved in the evaluation or the specific projects examined, thereby enriching the collective understanding and applicability of the research outcomes.

\section{Results}
\label{sec:Results}

In this section, we group and report the results according to the three research questions of this study.

\subsection{\texorpdfstring{\ref{rq1}}. Engineering practices when working with
requirements with regulation as a source}
Through interviews and meetings with practitioners from the studied cases, we identified activities that are the direct consequences of developing software systems with regulatory requirements. These activities are grouped and summarized into the following five themes.

\paragraph*{Understanding and Interpretation} In some cases, engineers must first understand and interpret the regulatory requirements, which often involve translating legal jargon into technical specifications. \textit{itestra} manages a diverse portfolio of clients and initiatives across various regulatory sectors. These sectors are characterized by varying degrees of complexity and specificity in their regulatory frameworks. For example, in cases A, B, and C the business analysts from \textit{itestra} are involved in the process of deriving requirements from regulations and interpreting them. The interviewee for case B describes this as ``\textit{....so sometimes there's also a business analyst involved who looks at this... So we kind of share the business analyst analysis part, with the customer who we deal with.}'' In contrast to case D, where \textit{itestra} receives the requirements specification document with the interpreted regulations. ``\textit{... you never get the regulation itself \footnote{This doesn't imply any concealment or withholding of information by the parties involved. Instead, it indicates that, typically, there's no necessity for the development team to engage with the regulatory texts themselves directly.}. You always get the specifications. The customer had their domain expert come up with the requirements.}'' Furthermore, the interviewee observed that ``\textit{Usually that word document quotes some of the legislation. As background information, but then states what actually needs to be implemented.}''

\paragraph*{Creation of Supplementary Documentation and Artifacts} Word and PDF documents, created by both domain experts from the client and business analysts from \textit{itestra}, serve as the main artifacts for capturing requirements specifications. For example, this includes Word or annotated PDF documents that capture modifications to the previous year's tax certificate for the given tax year.
However, Excel sheets and additional documents are created to detail the requirements and sometimes provide examples from regulations. The interviewee in case D stated that 
``\textit{... the main thing is in the Word document, sometimes accompanied with some Excel sheets showing examples}''.

\paragraph*{Collaboration with Domain Experts} 
Regular interaction with domain experts from the client ensures that the technical team at \textit{itestra} fully grasps the regulatory requirements. This communication occurs through various modes, including Excel files or Confluence pages, which list necessary modifications and serve as a basis for development and testing. The team engages in direct interactions, including meetings and stakeholder discussions. Information and clarifications are shared via Confluence pages, with notes taken during meetings. Project management utilizes a Kanban board, and while direct conversations were common, current communication primarily occurs through JIRA tickets, Word documents, and occasional phone calls.
Particularly in cases A, B, and D, \textit{itestra}'s analysts play a role in interpreting regulations to derive software requirements, whereas in case C, the client provides \textit{itestra} with pre-interpreted requirements. This distinction highlights the varied involvement of \textit{itestra}'s team across projects.
\paragraph*{Testing for Compliance} Rigorous testing, including regression testing to ensure that new changes do not disrupt existing functionalities, is critical to maintaining compliance with regulations. The interviewee in Case D stated that ``\textit{checking what actually changed or what differences are there between different tariffs variants... Testing and in particular regression testing is a huge topic.}'' 

\paragraph*{Feedback and Certification} In some cases, certification from external auditing bodies is required, which involves running test cases and reviewing results and processes to ensure compliance. For instance, cases A, C, and D needed some form of external auditors. The interviewee in Case C stated that ``\textit{... there was one auditing organization... They had to approve the general process}''.

\begin{tcolorbox} [colback=black!5!white,breakable,colframe=black!75!black,title=Highlights:]
Diverse practices recognized in managing regulatory requirements in software development focus on interpretation of regulations, documentation creation, and collaboration with domain experts. It also highlights the importance of rigorous testing for compliance and the necessity for external certification in some cases.
\end{tcolorbox}

\subsection{\texorpdfstring{\ref{rq2}}. Challenges encountered by software engineering teams when working with requirements with regulations as a source}

\paragraph*{Interpretation Difficulty} Regulations are often complex and written in legal language, making it difficult for engineers and even domain experts to interpret them accurately for implementation. In two of the cases, the analyst described it as ``\textit{I think the biggest challenge is to interpret the law. Interpret the text that is the basis for the requirements... I don't know, map it to what we're doing in the system?\,}'' Also, they stated: ``\textit{I think the challenges with the regulation X [anonymized] is that it's,... not a very specific requirement.}''
\paragraph*{Communication Gaps} Facilitating the same understanding and interpretation of legal concepts between domain experts and developers is a key challenge. 
Excel files and Confluence pages are usually used to align both parties on the requirements. However, the limited information available beyond the regulation and the lack of feedback loops and communication with lawmakers make it challenging to comply with the regulations. One interviewee mentioned: ``\textit{I think the biggest challenge is the limited information you have. So you have this communication handbook. And then other than that, there's practically nothing.}'' 
\paragraph*{Late Changes} Regulatory documents and their updates can arrive late in the development process, causing a rush to implement changes and verify compliance. In one case, it was mentioned that they try to foresee the future to implement the correct specification on time (e.g. generating the tax certificate). In an interview, it was mentioned ``\textit{That's also one challenge. I think the final version of this always, arrives pretty late in the process... It can be hard or even impossible to recalculate the correct values... So this lag also poses a challenge. I say, have to work well in advance and try to foresee the future and the regulations and everything.}'' 
\paragraph*{Testing and Verification} Ensuring that the software complies with the regulations requires thorough testing, which is time-consuming and often lacks automated processes. This is especially true when dealing with frequent updates to regulations. For example, in case B the interviewee stated ``\textit{CRS, I think is updated at least once a year because they have a list of countries that are included in this that are to be included in this report. ... So this is a list of four or five countries that always change the CRS.}'' 
\paragraph*{Documentation} There is a need for accurate documentation that reflects the regulatory requirements, changes made, and justifications for certain design decisions, which can be subject to audits. Maintaining accurate and up-to-date documentation that can be traced back to specific regulatory requirements is an ongoing challenge that needs attention. One interviewee stated: ``\textit{So you must be able to explain ten years later why this ... was put in this location and its complicated calculations in everything. And I think in this context, there was a requirement that you needed to be able to show documentation that fulfills certain regulatory requirements.}'' 
\paragraph*{Change Impact Analysis} Understanding and adapting to regulatory changes in software systems is difficult. It involves a thorough analysis to grasp how new regulations affect existing source code and systems. This process requires comparing previous and current regulations, pinpointing the differences, and then making necessary updates in the software. This involves identifying where the previous version of the regulation was implemented in the code so that those specific parts can be modified accordingly.
Such a task demands a methodical approach to ensure the software complies with evolving legislative changes. Figure~\ref{map} highlights this challenge, and it is significant as it demands deep technical understanding and a keen awareness of legal requirements. One interviewee commented ``\textit{I would say that is also a challenge, at least a minor challenge, to make sure that if you have a similar change several times, that it's done in the same way and that you don't always forget one of the things you're supposed to do or well, do it in a different way every time.}''

\begin{tcolorbox}[colback=black!5!white,breakable,colframe=black!75!black,title=Highlights:]
Identified challenges in integrating regulatory requirements into software development are the complexity of legal language, communication gaps between domain experts and developers, the impact of late regulatory changes, the necessity for comprehensive testing and verification, the importance of maintaining detailed documentation, and the difficulties in analyzing the impact of regulatory changes on existing systems.
\end{tcolorbox}

\subsection{\texorpdfstring{\ref{rq3}}. Potential of tool-supported approaches to address some of the identified challenges}
Based on the transcripts, tool-supported approaches for managing regulatory requirements show early promise, with several areas identified for potential benefits:

\paragraph*{Document Comparison} Automating the process of comparing new regulatory documents with previous versions to highlight changes and updates that need attention. An ideal tool for this task would possess "intelligence" beyond simple page-by-page comparison, capable of discerning alterations in phrasing and changes in semantics. Such a tool would significantly enhance the ability to identify where regulations have substantively changed, facilitating a more efficient update process in the software to address these modifications accurately. The interviewee in Case B mentioned ``\textit{I see a point of improvement there. If you could just have a process that you get the automatic update of this document. Accept these changes and then your lists in your software are automatically updated}''.

\paragraph*{Test Case Generation} Leveraging automated tools for generating test cases from requirements ensures comprehensive coverage of all new regulatory mandates. This approach becomes particularly valuable when regulations are frequently updated, as manual testing processes are both time-consuming and often lack the agility needed to adapt to these changes swiftly. This scenario underscores the necessity for automated testing solutions that can efficiently adjust to and accommodate these regular updates, ensuring ongoing compliance with evolving regulatory landscapes.

\paragraph*{Update Notifications} Creating an automated system that notifies developers of regulation changes that may affect their project, ensuring timely updates. One interviewee noted: ``\textit{an automated diff and maybe even add the new parts or the modifications automatically to our template.}'' and ``\textit{it's imaginable that the tools tell me well, last time there was a change that looked similar. Those were the code locations that you changed}''.

\paragraph*{Traceability} Implementing tools that provide traceability from requirements to implementation, making it easier to manage changes and their impact. One interviewee mentioned the following about automation benefits ``\textit{You need to know which parts of the system should be changed and how. But this is something very specific to the requirement.}''

\begin{tcolorbox}[colback=black!5!white,breakable,colframe=black!75!black,title=Highlights:]
Key identified areas suitable for tool-supported approaches include document comparison to identify regulatory changes, automated generation of test cases from requirements, update notifications to alert developers of regulatory changes, and tools for traceability to simplify the management of changes.
\end{tcolorbox}

\section{Discussion}
\label{sec:Discussion}

In this section, we critically discuss and compare our results with respect to the analyzed software practices, challenges, and potential enhancements with the related works.

\subsection{Engineering practices (\texorpdfstring{\ref{rq1}}))}
The first research question addressed the engineering practices for handling requirements sourced from regulations. The primary tools identified were conventional documentation like Word or PDF files, Excel spreadsheets, and digital collaboration platforms like Confluence. These are the main communication channels between domain experts and developers, ensuring both parties are aligned on the same topics. However, there is a lack of structure for the involvement of domain experts. Klymenko et. al.~\cite{b4} made similar observations and tried to address this issue in their paper. It seems that in cases where non-legal experts need to interpret the regulations, some form of training and education will be beneficial. The interviews do not indicate the significant involvement of external legal experts or auditors in the regular process, suggesting that most stakeholders are internal to \textit{itestra} and the customer company. One interesting finding is the lack of tool support and automation in the engineering processes. Particularly, while the regulations are well written in natural language, we expected some form of NLP-enabled approaches. This aligns well with Klymenko et al.'s findings on privacy engineers' unawareness of such tools \cite{b3}. 

\subsection{Challenges (\texorpdfstring{\ref{rq2}}))}
The identified challenges can be categorized into two broader groups:
\begin{itemize}
    \item \textit{Interpretative and Communication Challenges} \label{ch1}
    
    This group focuses on the challenges that arise from the need for expert interpretation, understanding of complex legal language, and effective communication between different stakeholders. It includes aspects where human judgment and expertise are crucial. This includes the difficulty in interpreting complex regulations and the communication gaps between domain experts and developers. It encapsulates the struggles of comprehending, interpreting, and translating regulations into actionable software requirements.
    \item \textit{Technical and Process-Oriented Challenges}\label{ch2}
    
    This group covers the challenges of managing the engineering process in response to regulatory changes. It involves aspects where automation, tool support, and structured methodologies can play a significant role. It includes dealing with late regulatory changes, ensuring thorough testing and verification for compliance, and maintaining detailed and up-to-date documentation to streamline these processes. This group reflects the operational difficulties in adapting to and staying compliant with evolving regulations.
\end{itemize}

Usman et al.'s~\cite{b5} industrial case study, which centers on compliance checking and analysis, shares some commonalities with our findings, particularly regarding challenges. Their study identifies a challenge category named "requirements specification-related challenges," which resonates with our initial two identified challenges. This category encompasses the complexities of interpreting compliance requirements for specific products and the varying understandings of these requirements, mirroring issues we have also observed in our research. Additionally, Usman et al.'s study highlights "Process-related challenges" in compliance checking and analysis, which corroborates our latter two identified challenges. These challenges include managing changes in compliance requirements and the absence of automation, paralleling issues we've identified in managing regulatory updates and the need for more automated processes in our own study. The challenge of interpreting regulatory requirements is a well-documented issue in the literature. For instance, regulations have been characterized as unclear~\cite{b6, b7}, as well as verbose, abstract, vague, and ambiguous~\cite{b8}. These descriptors highlight the inherent difficulties in deciphering regulations' exact meaning and application in practical scenarios. The communication gap challenge is closely associated with the need for domain expertise in processing and implementing regulations. Moyon et al.~\cite{b9} developed a pipeline to bridge the communication between security experts, and dev and ops teams. This was achieved by integrating security standards directly into DevOps pipelines, allowing for a seamless blend of security practices with the existing workflows of development and operations. In a context like \textit{itestra}'s, where various projects are subject to a wide range of regulations that not only differ significantly but also frequently update, this approach appears less viable. Similar challenges to mentioned late changes have been reported in other studies. Research from Granlund et al.~\cite{b10} and Wagner et al.~\cite{b11} highlights the conflict between the ongoing regulatory compliance demands and the need for continuous delivery and maintenance in software development. 
The testing and verification challenge, particularly the aspect of changing regulations over time, is also discussed by Hjerppe et al.~\cite{b12}. This study identifies the evolving nature of regulations as a key challenge in ensuring strong compliance assurance. Similar to the challenges of documentation and change impact analysis, Wagner et al.~\cite{b11} highlight issues with compliance-related documentation, noting its absence as a frequent problem. Concurrently, Ozcan-Top et al.~\cite{b13} emphasize the difficulties in tracking changes in requirements, underscoring the challenge of maintaining current and accurate documentation in the face of evolving regulatory landscapes. In some domains, such as automotive and embedded software engineering, there are established approaches for traceability. Maro et al.~\cite{b23} explore the complexities and propose solutions for software traceability within the automotive industry. The paper identifies key challenges, such as the vast scale of automotive software, the need for integrating multiple systems and standards, and the dynamic nature of automotive software development. They discuss various approaches and tools developed to enhance traceability, emphasizing the importance of robust traceability frameworks to ensure the reliability and safety of automotive software. 
Finally, Kellogg et al.~\cite{b14} explore the concept of "continuous compliance" within software engineering, focusing on the challenge of maintaining regulatory compliance throughout the development lifecycle, especially in agile and fast-paced environments. They discuss the gap in current practices where automation is seldom utilized to seamlessly ensure compliance is integrated into the development process. The study suggests that incorporating automated tools and methods to verify compliance at various stages of software development could significantly reduce the burden on developers to manually ensure adherence to regulations, thus enhancing efficiency and reliability in meeting compliance requirements.

\subsection{Potential for tool support (\texorpdfstring{\ref{rq3}}))}

Tool-supported approaches have significant potential for addressing software engineering challenges related to regulatory compliance. Through tool support, repetitive tasks can be streamlined for efficiency, and consistency across systems can also be maintained rigorously. For example, in the context of a payroll system, automation could adaptively manage and implement new types of employee supplements as regulatory requirements change. 
It also enables engineers to efficiently discern which system components are impacted by new regulations. An advanced approach involves utilizing automated tools for comparing previous and new regulatory documents to spotlight modifications and integrate required changes into the system's design directly. Additionally, an intermediate step might include checking code segments previously modified for compliance—using indicators such as JIRA ticket numbers in commit messages—to guide developers directly to the relevant sections for updates.
Deciding on changes to calculations or interpreting the deeper meaning of regulatory texts may still require human judgment. This resonates with the observations from the interviewees across the different cases, excluding case C, where tool support was not suggested as a remedy for the challenges identified (the challenges are mainly from group one \ref{ch1}). Contrarily, participants from three separate instances pinpointed areas such as document comparison, traceability, and update notifications where tool-supported approaches could significantly contribute to their operational efficiency. The mentioned potentials for improvement can turn into strategic insights and action plans. 

\subsection{Threats to validity}
In our multi-case study, we recognize several threats to validity that may influence the results and their interpretation:

\textit{Internal Validity}: The reliance on semi-structured interviews may lead to subjective interpretations of participants' responses. We attempted to mitigate this by carefully reviewing the interview questions by a senior researcher and sending the study results to participants for verification.

\textit{Construct Validity}: The potential for misunderstanding or miscommunication between the researchers and practitioners during interviews may affect the accuracy of the data collected. We have tried to reduce this risk by preparing detailed interview guides and engaging in follow-up discussions. 

\textit{External Validity}: Our findings are based on a single company's experiences, which may not be generalizable to other contexts or industries.
However, our aim was not to reveal universally valid claims and generalisations but to conduct an in-depth study with our partner company that considers practices and challenges in one selected domain. The specific nature of \textit{itestra’s} domain and the type of regulations they work with made this possible, although, of course, this may well limit the applicability of our results to other domains.

\textit{Reliability}: The reproducibility of this study may be affected by the fact that we cannot disclose all details of the cases due to confidentiality and the unique circumstances under which the study was generally conducted. Although we have documented our methodology thoroughly, the dynamic and complex nature of regulatory requirements could lead to different results if the study were to be replicated. 
The selection of the industry partner was influenced by an existing long-term collaboration, which may be well perceived as opportunistic but is nevertheless important considering the highly confidential environment we could jointly analyze. We further believe the chosen company setting is well-suited and representative for our research focus. We have gathered insights from various stakeholders, giving us confidence in the comprehensiveness and reliability of the collected perspectives and opinions.

Future research should address these threats by expanding the scope of study to include a broader range of companies and regulatory contexts and by employing quantitative methods to complement the qualitative insights gained from this study.

\section{Conclusion and Future Work}
\label{sec:Conclusion}

Our multi-case study within \textit{itestra} has provided substantial insights into the complexities of deriving software requirements from regulatory sources. We have discussed the practical challenges software engineers face and the potential for tool-supported approaches to mitigate these difficulties. In particular, interpretation difficulty, communication gaps, late changes, testing and verification, documentation, and change impact analysis have been identified as significant hurdles.

The study has also highlighted the necessity for a sensible approach to requirement engineering in regulated domains, particularly emphasizing the fintech sector. While tool support and automation seem promising, particularly in document comparison and update notifications, its application is still limited when it comes to interpreting regulations and understanding the particularities of legal language. Here, one promising direction is the current advances in Natural Language Processing (especially the use of large language models) to support the interpretation (and translation) of legal language.

For future work, we see the development of sophisticated tools that can aid in interpreting regulations, offer more robust traceability, and provide automated updates to documentation of particular importance. Additionally, further research should focus on enhancing communication channels between domain experts and developers to streamline the process of converting regulations into actionable software requirements.

As regulations continue to evolve, the need for agility and precision in software development will only increase. One hope we associate with our contribution is to provide a stepping stone towards addressing these challenges, with the ultimate goal of supporting the creation of software engineering environments that are both compliant and efficient.

\subsection*{Data Availability}
The interview study protocol is available at \href{https://doi.org/10.5281/zenodo.10640987}{zenodo} \footnote{https://doi.org/10.5281/zenodo.10640987}
\section*{Acknowledgment}
We thank our supportive long-term partner \textit{itestra} and all practitioners involved in our study for making this study possible and providing such valuable insights. This work was further supported by the KKS foundation through the S.E.R.T. Research Profile project at Blekinge Institute of Technology.

\end{document}